\documentclass[prd,twocolumn,showpacs,superscriptaddress,floatfix,longbibliography]{revtex4-1}
\usepackage{graphicx,amsfonts,amssymb,amsmath,xspace,dsfont}
\usepackage[colorlinks=true,citecolor=green,linkcolor=red,urlcolor=blue]{hyperref}

\usepackage{color}
\definecolor{g}{rgb}{.1,0.4,.1} % {.0,0.7,.5}
\definecolor{b}{rgb}{0,0.2,1}
\definecolor{rouge}{rgb}{0.82,0.,0.}
\definecolor{vert}{rgb}{0.,0.82,0.}
\definecolor{orange}{rgb}{1,0.5,0.}
\definecolor{bleu}{rgb}{0.,0.,0.82}
\definecolor{m}{rgb}{0.82,0.,0.82}
\definecolor{vert2}{rgb}{0.,0.5,0.}
\definecolor{rougeclair}{rgb}{1.0,0.7,0.7}

\begin{document}

\title{Bound states in string nets}

\author{Marc Daniel Schulz}
\email{mdschulz@umn.edu}
\affiliation{Department of Physics and Astronomy, University of Minnesota, Minneapolis, Minnesota 55455, USA}
\author{S\'{e}bastien Dusuel}
\email{sdusuel@gmail.com}
\affiliation{Lyc\'ee Saint-Louis, 44 Boulevard Saint-Michel, 75006 Paris, France}
\author{Julien Vidal}
\email{vidal@lptmc.jussieu.fr}
\affiliation{Laboratoire de Physique Th\'eorique de la Mati\`ere Condens\'ee,
CNRS UMR 7600, Universit\'e Pierre et Marie Curie, 4 Place Jussieu, 75252
Paris Cedex 05, France}

\begin{abstract}
We discuss the emergence of bound states in the low-energy spectrum of the string-net Hamiltonian in the presence of a string tension. In the ladder geometry, we show that a single bound state arises either for a finite tension or in the zero-tension limit depending on the theory considered. In the latter case, we perturbatively compute  the binding energy as a function of the total quantum dimension. We also address this issue in the honeycomb lattice where the number of bound states in the topological phase depends on the total quantum dimension. Finally, the internal structure of these bound states is analyzed in the zero-tension limit. 
\end{abstract}


\maketitle
%%%%%%%%%%%%%%%%%%%%%%%%%%%%%%%%%%%%%%%%%%%%%%
%%%%%%%%%%%%%%%%%%%%%%%%%%%%%%%%%%%%%%%%%%%%%%
%%%%%%%%%%%%%%%%%%%%%%%%%%%%%%%%%%%%%%%%%%%%%%

%
%
%%%%%%%%%%%%%%
\section{Introduction}
%%%%%%%%%%%%%%
%
%

Interactions between elementary particles are known to be responsible for the formation of bound states of matter. In particle physics, the most celebrated examples are baryons and mesons stemming from the strong interaction between quarks mediated by gluons. Bound states often have features very different from those of their constituents. For instance, Cooper pairs that originate from the interaction between electrons and phonons in solids give rise to superconductivity whose existence is due to the bosonic nature of these quasiparticles. Understanding formation mechanisms of bound states as well as their properties is thus of primary interest especially when the building blocks are exotic excitations which are neither bosons nor fermions.  Such objects, known as anyons~\cite{Leinaas77, Wilczek82}, have drawn much attention because of their potential use for topological quantum computation \cite{Kitaev03,Freedman03,Preskill_HP,Nayak08,Wang_book}. Although anyons have not yet been unambiguously detected in experiments, the most promising candidates are fractional quantum Hall states~\cite{Moore91,Wen91}. The concept of anyon is intimately related to the notion of topological quantum order whose description requires us to go beyond the Landau symmetry-breaking theory (see Ref.~\cite{Wen13} for a review). Recently, substantial progress has been made to classify topological phases, especially for $(2+1)$-dimensional systems~\cite{Rowell09,Wen16,Schoutens16} but a systematic way to microscopically engineer them is still lacking. A major step in this direction has been taken by Levin and Wen~\cite{Levin05} through the so-called string-net model that gives access to a broad class of topological phases~\cite{Lin14,Lake16}. 

The goal of the present paper is to study the formation of bound states induced by the introduction of a string tension in the string-net model. A string tension is known to break the topological phase when it is sufficiently large as already discussed in several cases~\cite{Gils09_1,Gils09_3,Ardonne11,Burnell11_2,Schulz13,Schulz14,Schulz15,Dusuel15,Schulz16}. Here, we focus on the quasiparticle spectrum in the topological phase and we show that, in most cases, bound states are the low-energy excitations. We compute the binding energy in the weak-tension limit for any anyon theory in the two-rung ladder and in the honeycomb lattice. In the latter case, we find that the number of bound states depends on the total quantum dimension.  Finally, we investigate the internal structure, i.e., the particle content of these bound states. In the one-dimensional case, we find that the contribution of each anyon type to the bound states is always proportional to their quantum dimension whereas this symmetry does not hold for large tension in the two-dimensional case.

%
%
%%%%%%%%%%%%%%%%%%%%%%%
%%%%%%%%%%%%%%%%%%%%%%%
\section{Hilbert space and string-net model}
%%%%%%%%%%%%%%%%%%%%%%%
%%%%%%%%%%%%%%%%%%%%%%%
%
%
In the string-net model introduced in Ref.~\cite{Levin05}, microscopic degrees of freedom are defined on the links of a trivalent graph and can be in $N$ different states where $N$ is given by the input theory of interest. Mathematically, this input theory is described by a unitary modular tensor category which, at this stage, can be roughly considered as a set of objects obeying a set of rules (see for instance Refs.~\cite{Rowell09,Wang_book} for details). The Hilbert space of the string-net model is spanned by all link configurations satisfying the so-called branching rules at each vertices. These branching rules stem from the fusion rules of the input theory: at a given trivalent vertex, if two links are in the state $a$ and $b$, the third one must be in a state $c$ that belongs to the fusion product $a$ and $b$. Violations of these branching rules correspond to charge excitations that are not considered here.

The Hamiltonian of the string-net model is given by a sum of local commuting projectors acting on plaquettes 
%
%
%%%%%%%%%%%%%%%%
\begin{equation}
H_0=-\sum_p B_p.
\label{eq:ham_LW}
\end{equation}
%%%%%%%%%%%%%%%%
%
%
The explicit action of $B_p$ on a link configuration is given in Ref.~\cite{Levin05}. By definition, we call an eigenstate of $B_p$ with eigenvalue  $1$ $(0)$ a state without (with) a flux in the plaquette $p$. The ground-state manifold of $H_0$ is thus spanned by all states without flux in plaquettes. The dimension $\mathcal{D}_0$ of this manifold, also known as the topological degeneracy, depends on the system. On a compact surface of genus $g$, $\mathcal{D}_0=N^{2g}$ whereas $\mathcal{D}_0=N$ for a ladder with periodic boundary conditions.

By acting with a local operator on any ground state, excitations can only be created by pairs \footnote{For non-Abelian theories and for $g>0$, one may generate single-flux excitations by using nonlocal operators.} 
The degeneracy $\mathcal{D}_k$ of the excited states with $k$ fluxes depends both on the surface topology and on the fusion rules of the theory considered (see Refs.~\cite{Schulz13,Schulz14,Schulz15} for various examples). The corresponding excitation energy $\Delta_k=E_k-E_0=k$  does not depend on the position of the fluxes (deconfined phase). 

The main focus of this work is to analyze the excitation spectrum in the presence of a perturbation that lifts this degeneracy and effectively provides quantum dynamics to fluxes.  To this aim, let us consider the operator
%
%
%%%%%%%%%%%%%%%%
\begin{equation}
V=-\sum_l T_l,
\label{eq:ham_pert}
\end{equation}
%%%%%%%%%%%%%%%%
%
%
where $T_l$ is a projector which is diagonal in the link basis. If a link $l$ is in a state $a$ then $T_l=\delta_{a,0}$ where $0$ denotes the trivial state. The perturbation $V$ is  local and it does not commute with $H_0$. In the eigenbasis of $H_0$ (flux basis), $T_l$ acts effectively as a creation/destruction operator but also as an interaction/transmutation term between fluxes. From that respect, $V$ plays the role of a string tension that eventually induces a transition towards a confined phase~\cite{Gils09_1,Gils09_3,Ardonne11,Burnell11_2,Schulz13,Schulz14,Schulz15,Dusuel15}.

In the following, we consider the Hamiltonian \mbox{$H=H_0+ \kappa \:V$} and, for simplicity, we restrict our discussion to the \mbox{case $\kappa \geqslant 0$}. For the ${\mathbb Z}_N$ theory, this model can be mapped onto the transverse-field Potts model where fluxes are microscopic degrees of \mbox{freedom~\cite{Burnell11_2}}.  Since $V$ is a sum of local terms, it does not perturbatively change  the flux through noncontractible loops that may exist for some surface topologies. This property which is the source of the so-called topological protection allows one to  study each topological sector separately. Below, we focus on the trivial sector in which elementary excitations for $\kappa=0$ are all possible pairs of fluxes $(a,\bar a)$ fusing to the vacuum ($\bar a$ denotes the dual of $a$). 
For $\kappa \ll 1$, excitations can be seen as quasiparticles (QPs) corresponding to ``dressed fluxes" in the usual Fermi liquid picture.

%
%
%%%%%%%%%%%%%%%%%%%%%%%
%%%%%%%%%%%%%%%%%%%%%%%
\section{The one-dimensional case}
%%%%%%%%%%%%%%%%%%%%%%%
%%%%%%%%%%%%%%%%%%%%%%%
%
%
Let us start our analysis by considering the string-net model defined on the two-leg ladder. As discussed in Refs.~\cite{Gils09_1,Gils09_3,Schulz15}, the string-net ladder in the presence of the perturbation $V$ undergoes a quantum phase transition at the self-dual point $\kappa=1$. At this point, the model can be exactly mapped onto the $XXZ$ chain with an anisotropy para\-meter $D/2$ where $D$ is the total quantum dimension of the theory. This problem is also equivalent to the \mbox{$q$-state} Potts model  with $q=D^2$. Consequently, if $D\leqslant 2$ the self-dual point is a critical point~\cite{Gils09_1,Gils09_3,Ardonne11} whereas for $D>2$ it corresponds to a first-order transition point~\cite{Schulz15}. 

At order 1 in perturbation, in the limit $\kappa \ll 1$, one can exactly compute the 2QP spectrum for any theory. Therefore, one has to compute the matrix elements of $V$ in the 2QP subspace. In the trivial sector, let us denote $|p^{a}_{1},p^{\bar a}_{2}\rangle$ the unique state with a flux $a$ ($\bar a$) in the plaquette \mbox{$p_1$ ($p_2$)}. If $p_1$ and $p_2$ are not adjacent, $V$ cannot change the flavor of the fluxes and acts as a single-flux hopping term. Indeed, one has
%
%
%%%%%%%%%%%%%%%%
\begin{equation}
\langle p^{a}_{3},p^{\bar a}_{4}|V|p^{a}_{1},p^{\bar a}_{2}\rangle=-\frac{1}{D^2} \Big(N_{\rm l}\delta_{p_3}^{p_1}\delta_{p_4}^{p_2}+\chi_{p_3}^{p_1} \delta_{p_4}^{p_2}+\chi_{p_4}^{p_2}\delta_{p_3}^{p_1}\Big),
\label{eq:V_hopping}
\end{equation}
%%%%%%%%%%%%%%%%
%
%
where $\chi_{p_j}^{p_i}=1$ if $p_i$ and $p_j$ are adjacent plaquettes and 0 otherwise,  $N_{\rm l}$ is the total number of links, and $\delta_{p_j}^{p_i}$ is the usual Kronecker symbol. By contrast, if $p_1$ and $p_2$ are adjacent, $V$ acts as an interaction term that may also transmute a pair $(a,\bar a)$ into a pair $(b,\bar b)$. More precisely,
%
%
%%%%%%%%%%%%%%%%
\begin{equation}
\langle p^{b}_{1},p^{\bar b}_{2}|V|p^{a}_{1},p^{\bar a}_{2}\rangle=-\frac{1}{D^2} \Bigg(N_{\rm l}\: \delta_{b}^{a}+\sum_{c \in a \times b} d_c \Bigg),
\label{eq:V_interaction}
\end{equation}
%%%%%%%%%%%%%%%%
%
%
where the sum runs over all nontrivial fluxes $c$ that belong to the fusion product $a\times b$, and $d_c$ denotes the quantum dimension of the flux $c$. 
Furthermore,  for any ground state $|0\rangle$ of $H_0$, one has $\langle 0|V|0\rangle=-N_{\rm l}/D^2$.

%
%%%%%%%%%%%%%%%%%%%%%%%
\begin{figure}[t]
\includegraphics[width=0.9\columnwidth]{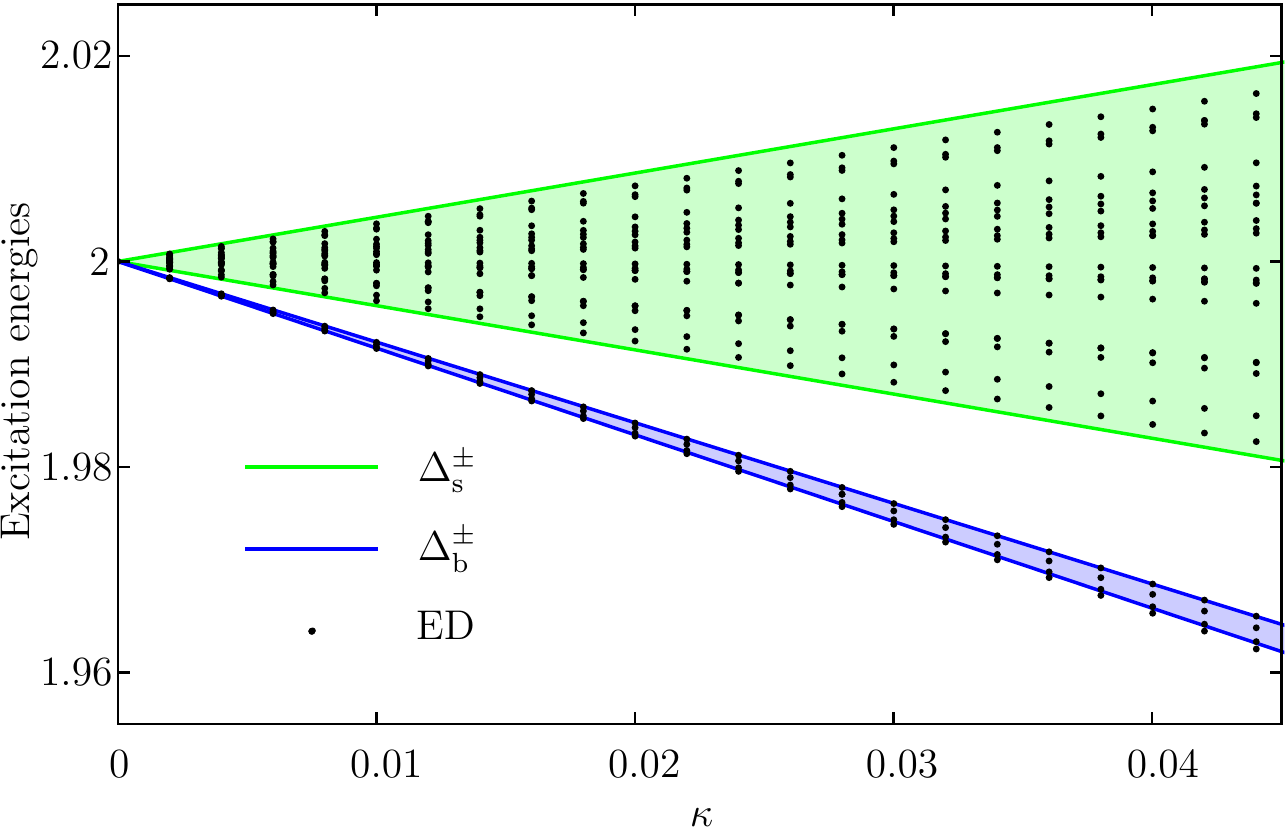}
\caption{Low-energy excitation spectrum  for the $(A_1,5)_{1/2}$ theory~\cite{Schulz15} on a ladder with seven plaquettes and periodic boundary conditions. The scattering-states continuum (green) and the bound-state continuum (blue) computed perturbatively at order 1 for $\kappa \ll 1$ are compared with exact diagonalization (ED) results. }
\label{fig:bound_states}
\end{figure}
%%%%%%%%%%%%%%%%%%%%%%%
%

In the thermodynamical limit, one can analytically diagonalize $V$ in the 2QP subspace. For any $D$, the 2QP excitation spectrum consists of a continuum of scattering states. As can be straightforwardly derived from Eq.~(\ref{eq:V_hopping}), the boundaries of this continuum are given by

%
%
%%%%%%%%%%%%%%%%
\begin{equation}
\Delta^{\rm \pm}_{\rm s}=2 \pm \kappa \frac{4 }{D^2} +\mathcal{O}\big(\kappa^2 \big).
\label{eq:2-QP_cont_gap_1D}
\end{equation}
%%%%%%%%%%%%%%%%
%
%
More interestingly, when the total quantum dimension $D>2$, one finds a continuum associated to a low-energy bound state whose boundaries are given by
%
%
%%%%%%%%%%%%%%%%
\begin{eqnarray}
\Delta^+_{\rm b}&=&2-\kappa \frac{D^2-2}{D^2} +\mathcal{O}\big(\kappa^2\big), \label{eq:2-QPgap_1D_1}\\
\Delta^-_{\rm b}&=&2-\kappa \frac{8-4 D^2+D^4}{D^2(D^2-2)} +\mathcal{O}\big(\kappa^2\big). \label{eq:2-QPgap_1D_2}
\end{eqnarray}
%%%%%%%%%%%%%%%%
%
% 
Figure \ref{fig:bound_states} displays  a comparison between these results valid in the thermodynamical limit and exact diagonalization data obtained for the theory $(A_1,5)_{1/2}$ for which $D=\frac{\sqrt{7}}{2 \sin(\pi/7)}$.

For $D \leqslant 2$, there is no bound state in the spectrum at order 1 and one thus needs to investigate higher-order perturbation theory. Indeed, the operator $V$ is similar to the one considered in Refs.~\cite{Dorey03_1,Dorey03_2} to perturb the critical $q$-state Potts model. Keeping in mind that our model corresponds to the case $q=D^2$,  we know that in the vicinity of $\kappa=1$, bound states arise for \mbox{$\sqrt{3}<D \leqslant 2$}~\cite{Dorey03_1,Dorey03_2}.  This concerns two sets of modular theories: the Fibonacci theory for which $D=\sqrt{1+\phi^2}$ (where \mbox{$\phi=\frac{1+\sqrt{5}}{2}$} is the golden mean) and all theories with $D=2$ (see  Ref.~\cite{Rowell09} for details). However, let us remind that our perturbative analysis is performed in the vicinity of $\kappa=0$ and one may wonder whether fingerprints of the bound states present near $\kappa=1$ can be observed. 

%
%
%%%%%%%%%%%%%%%%%%%%%%%
\begin{figure}[t]
\includegraphics[width=0.9\columnwidth]{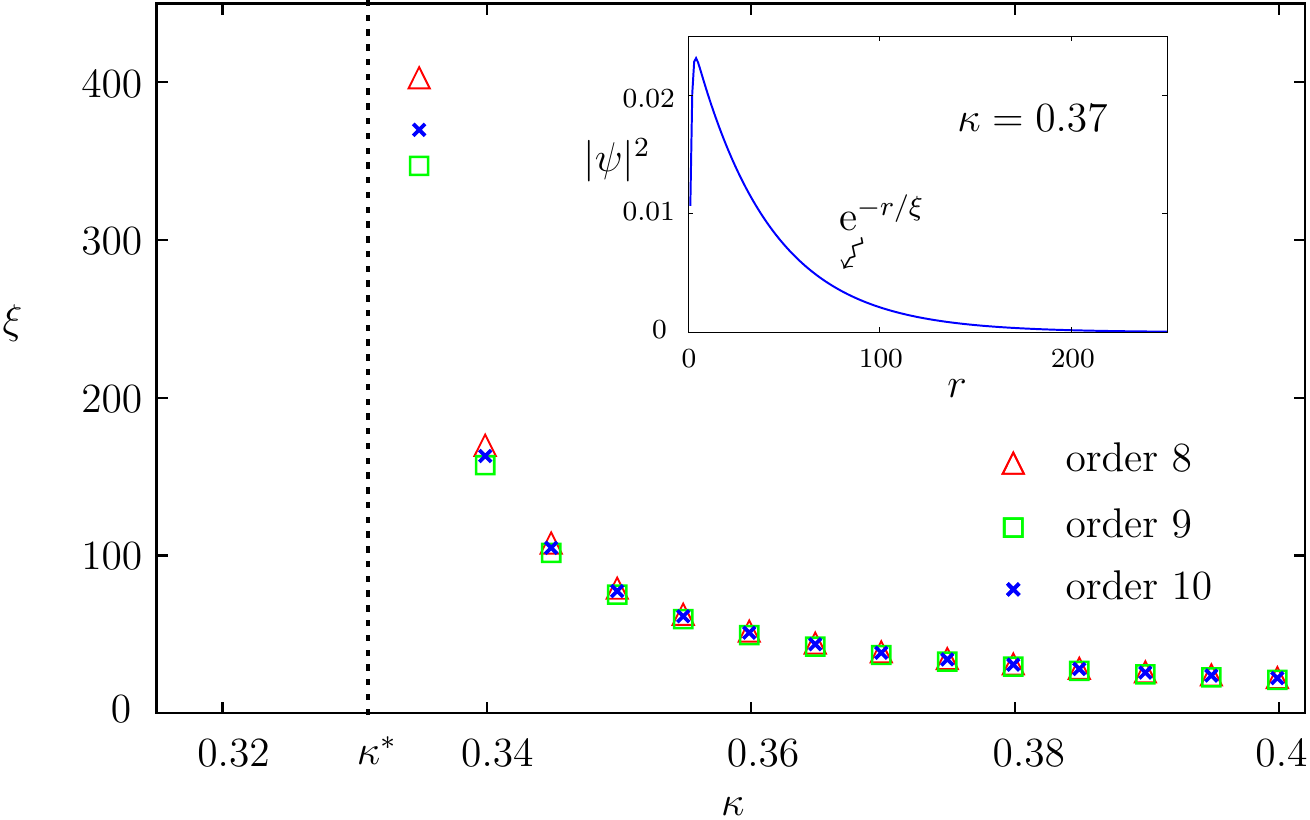}
\caption{Localization length $\xi$ as a function of $\kappa$ for the Fibonacci theory, measured in units of the distance between adjacent plaquettes. Inset: probability density of the (translation-invariant) lowest-energy bound state as a function of the distance between the two fluxes. For $\kappa=0.37$, at order 10, the binding energy is $\Delta_\mathrm{s}^- -\Delta_\mathrm{b}^- \simeq 4.82 \times 10^{-5}$, $|\psi |^2$ is maximum for $r=4$ and $\xi \simeq 39.84$.
}
\label{fig:Localization}
\end{figure}
%%%%%%%%%%%%%%%%%%%%%%%
%
%

To address this issue, we computed the 2QP spectrum up to order 10 in the limit $\kappa \ll 1$.  For all $D=2$ theories, we found one bound state in the whole range $\kappa \in [0,1]$, but the binding energy is of order 3 contrary to the case $D>2$  where it is of order 1 [see Eqs.~(\ref{eq:2-QP_cont_gap_1D})--(\ref{eq:2-QPgap_1D_2})]. More precisely, for $D=2$, one gets  
%
%
%%%%%%%%%%%%%%%%
\begin{eqnarray}
\Delta^-_{\rm s}&=&2- \kappa - \frac{3}{4}\kappa^2 + \frac{1}{16} \kappa^3+\mathcal{O}\big(\kappa^4\big),\\
\Delta^-_{\rm b}&=&2- \kappa - \frac{3}{4}\kappa^2 - \frac{1}{128} \kappa^3+\mathcal{O}\big(\kappa^4\big).
\label{eq:2-QP_D2}
\end{eqnarray}
%%%%%%%%%%%%%%%%
%
%
This bound state identifies with the state predicted in the perturbed four-state critical Potts model but the present perturbative  approach for $\kappa\ll 1$ does not allow us to recover the exact critical mass ratio which for $D=2$ is given by
 \mbox{$\displaystyle{\lim_{\kappa \rightarrow 1}} \Delta^-_{\rm b}/\Delta^-_{\rm s}=\sqrt{3}/2$}~\cite{Dorey03_1,Dorey03_2}.  
 
For the Fibonacci theory, we also find one bound state but only in the range \mbox{$\kappa \in [\kappa^*,1]$ where $\kappa^*=0.331(2)$}. Indeed, as can be seen in Fig.~\ref{fig:Localization}, the localization length of the bound state diverges at \mbox{$\kappa=\kappa^*$} and, for smaller values of $\kappa$, low-energy excited states are scattering states.  
Here again, we do not recover the exact critical mass ratio which, for \mbox{$D=\sqrt{1+\phi^2}$}, is given by \mbox{$\displaystyle{\lim_{\kappa \rightarrow 1}} \Delta^-_{\rm b}/\Delta^-_{\rm s}=(1+\sqrt{3})/2^{3/2}$}~\cite{Dorey03_1,Dorey03_2}.

%
%
%%%%%%%%%%%%%%%%%%%%%%%
\begin{figure}[t]
\includegraphics[width=0.9\columnwidth]{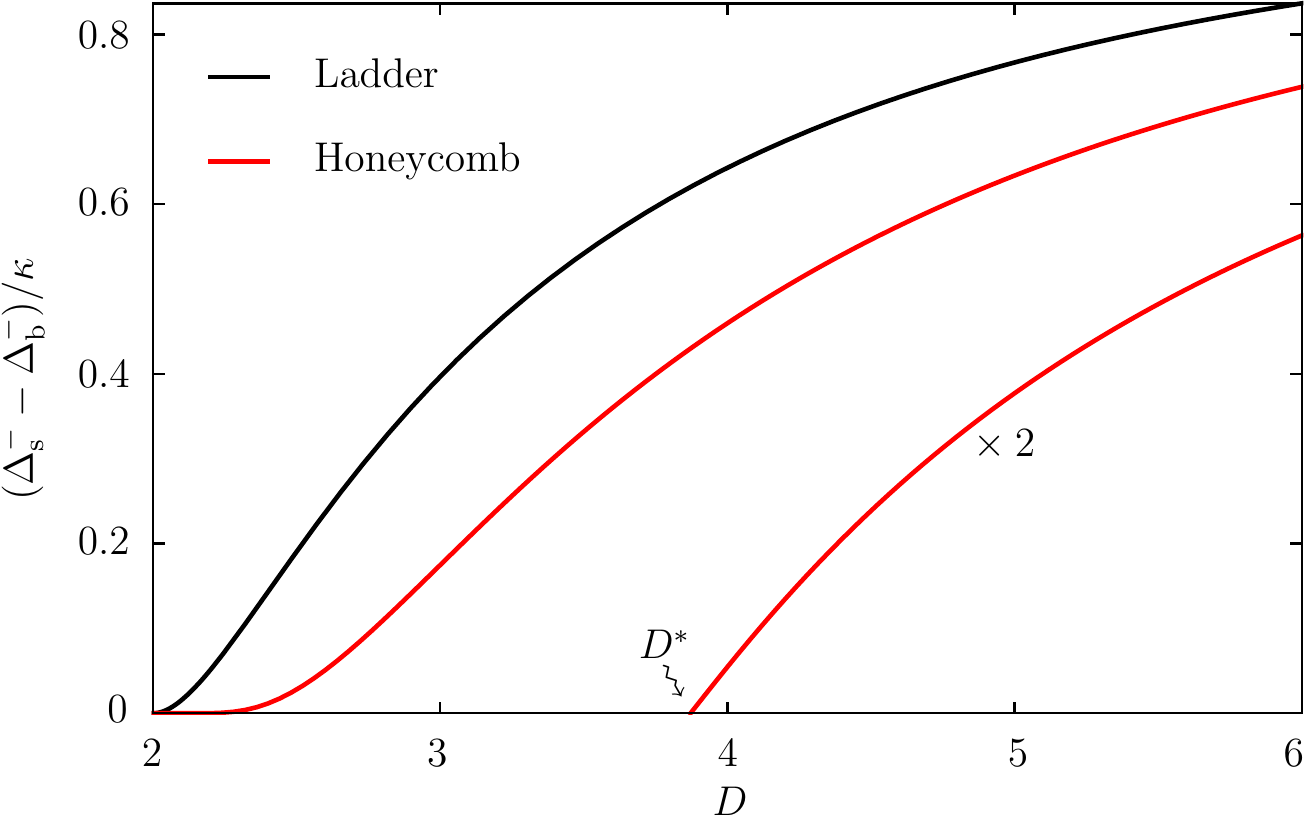}
\caption{Rescaled binding energies computed perturbatively at order 1 for $\kappa \ll1$ as a function of total quantum dimension for the ladder (black) and for the honeycomb lattice (red). In the latter case, two additional bound states emerge for $D \geqslant D^*$. For both geometries, one further has:  $\displaystyle{\lim_{D\rightarrow \infty}} (\Delta^{\rm -}_{\rm s}-\Delta^{\rm -}_{\rm b})/\kappa=1$.
}
\label{fig:Binding}
\end{figure}
%%%%%%%%%%%%%%%%%%%%%%%
%
%

For completeness, let us mention that we did not find any 2QP bound state in the ladder for $D \leqslant \sqrt{3}$, i.e., for semion ($D= \sqrt{2}$) and $\mathbb{Z}_3$ ($D= \sqrt{3}$) theories, in agreement with results from Refs.~\cite{Dorey03_1,Dorey03_2}.

%
%
%%%%%%%%%%%%%%%%%%%%%%%
%%%%%%%%%%%%%%%%%%%%%%%
\section{The two-dimensional case}
%%%%%%%%%%%%%%%%%%%%%%%
%%%%%%%%%%%%%%%%%%%%%%%
%
%

Let us now consider the honeycomb lattice which is the simplest trivalent two-dimensional system. Contrary to the one-dimensional case, we cannot provide an explicit form of the binding energy, even at order 1. Thus, results given thereafter have been obtained by numerically diagonalizing the effective 2QP Hamiltonian whose matrix elements have been computed perturbatively in the limit $\kappa \ll 1$ for each theory of interest. The maximum relative distance between particles considered here is $r_{\rm max}= 2^{11}$ (in units of the distance between adjacent plaquettes).

At order 1, matrix elements of $V$ in the 2QP subspace are still given by Eqs.~(\ref{eq:V_hopping}) and (\ref{eq:V_interaction}) but plaquettes now form a triangular lattice. At this order, the spectrum only depends on $D$ and three cases must be distinguished: \\\\

\noindent $\bullet$ $D\leqslant 2$: no low-energy bound state; \\

\noindent $\bullet$ $2 < D \leqslant D^*\simeq  3.87145$: one low-energy bound state; \\

\noindent $\bullet$ $D^*<D$: three low-energy bound states among which two have the same binding energy. \\

\noindent

The corresponding binding energies are displayed in Fig.~\ref{fig:Binding}.

At higher orders, the situation is different. Indeed, at order 2 and beyond, we also find one bound state for all theories with $D\leqslant 2$ except for semions ($D=\sqrt{2}$). This bound state emerges for a finite string tension $\kappa^*$ that depends on the theory. Unfortunately, in two dimensions, we cannot reach sufficiently high orders to obtain accurate va\-lues of $\kappa^*$. Let us simply mention that for the Ising theory, one gets $\kappa^* \simeq 0.047$ at order 2 (instead of $\kappa^*=0$ in the ladder). 

In the absence of bound states, scattering states are the relevant low-energy excitations and their condensation generates a continuous transition. Such a scenario occurs for the semion theory where a second-order transition is expected \mbox{at $\kappa_{\rm c} \simeq 0.2097$~\cite{He90}}. To obtain a first-order transition, bound states are thus required in the low-energy spectrum as exemplified in the  $\mathbb{Z}_3$ theory where such a transition occurs \mbox{at $\kappa_{\rm c} \simeq 0.2466$~\cite{Hamer92}}. However, although necessary, we emphasize that the existence of bound states is not a sufficient condition to have a first-order transition as we have seen in the ladder for \mbox{$\sqrt{3}<D\leqslant 2$}.

%
%
%%%%%%%%%%%%%%%%%%%%%%%
%%%%%%%%%%%%%%%%%%%%%%
\section{Internal structure of the bound states}
%%%%%%%%%%%%%%%%%%%%%%%
%%%%%%%%%%%%%%%%%%%%%%%
%
%

For theories with $N>2$, an important issue concerns the particle content of the bound states. At order 1, in the limit $\kappa \ll1$, the 2QP spectrum is obtained by diagonalizing $V$ in the 2QP subspace.  The corresponding matrix elements given in Eqs.~(\ref{eq:V_hopping}) and (\ref{eq:V_interaction}) unveil a remarkable property of the eigenstates at this order. Indeed, the hopping term does not depend on the particle type and the interaction term simply depends on the fusion rules. Consequently, one can show that for any bound state $| \psi \rangle$ in the trivial flux sector one has:
%
%
%%%%%%%%%%%%%%%%%%%%%%%
\begin{equation}
\frac{\langle \psi |p^{a}_{i},p^{\bar a}_{j}\rangle}{\langle \psi |p^{b}_{i},p^{\bar b}_{j}\rangle}=\frac{d_a}{d_b}.
\label{eq:ratio}
\end{equation}
%%%%%%%%%%%%%%%%%%%%%%%
%
%
In other words, at order 1, bound states of $H$ are a superposition of all particle types with a weight proportional to the quantum dimension of the particles. In addition, in the ladder, we found that this structure is robust for all theories and all orders we studied. We thus conjecture that it results from a symmetry that we have not been able to evidence non perturbatively. However, in the honeycomb lattice, Eq.~(\ref{eq:ratio}) only holds up to \mbox{order 4}. At higher orders, nontrivial braiding between virtual states occur in the perturbation theory, single-flux hopping terms depends on the particle type, and the particle content depends on the order.

%
%
%%%%%%%%%%%%%%%%%%%%%%%
%%%%%%%%%%%%%%%%%%%%%%
\section{Perspectives} 
%%%%%%%%%%%%%%%%%%%%%%%
%%%%%%%%%%%%%%%%%%%%%%%
%
%
In this study, we have shown that a string tension could generate 2QP bound states in the string-net model. These bound states are bosonic since they are found in the trivial flux sector. However, we also found anyonic 2QP bound states in other flux sectors but their analysis is beyond the scope of the present work. 
Finally, we emphasize that these bound states are the relevant low-energy excitations of the system that would be directly observed in spectral densities for $\kappa \ll 1$ (see Ref.~\cite{Kamfor14} for a similar case in the toric code model in the presence of a magnetic field). 

We have argued that bound states have to be present in the excitation spectrum to build up a (nontrivial) first-order phase transition. In this case, when $\kappa$ increases, a cascade of level crossings associated to $n$QP ($n>2$) bound states is expected. However, as early realized in the Ising chain in a magnetic field, bound states may also be observed near critical points~\cite{Zamolodchikov89,Coldea10}.
A deeper understanding of the role played by the bound states in quantum phase transitions is clearly needed and we hope that the present work will stimulate further studies in this direction.

\acknowledgments

We thank B. Dou\c{c}ot and B. Estienne for fruitful and insightful discussions.

%\bibliography{/Users/vidal/Documents/TeX/Bibliotheques/bibliotheque4.bib}

\begin{thebibliography}{33}%
\makeatletter
\providecommand \@ifxundefined [1]{%
 \@ifx{#1\undefined}
}%
\providecommand \@ifnum [1]{%
 \ifnum #1\expandafter \@firstoftwo
 \else \expandafter \@secondoftwo
 \fi
}%
\providecommand \@ifx [1]{%
 \ifx #1\expandafter \@firstoftwo
 \else \expandafter \@secondoftwo
 \fi
}%
\providecommand \natexlab [1]{#1}%
\providecommand \enquote  [1]{``#1''}%
\providecommand \bibnamefont  [1]{#1}%
\providecommand \bibfnamefont [1]{#1}%
\providecommand \citenamefont [1]{#1}%
\providecommand \href@noop [0]{\@secondoftwo}%
\providecommand \href [0]{\begingroup \@sanitize@url \@href}%
\providecommand \@href[1]{\@@startlink{#1}\@@href}%
\providecommand \@@href[1]{\endgroup#1\@@endlink}%
\providecommand \@sanitize@url [0]{\catcode `\\12\catcode `\$12\catcode
  `\&12\catcode `\#12\catcode `\^12\catcode `\_12\catcode `\%12\relax}%
\providecommand \@@startlink[1]{}%
\providecommand \@@endlink[0]{}%
\providecommand \url  [0]{\begingroup\@sanitize@url \@url }%
\providecommand \@url [1]{\endgroup\@href {#1}{\urlprefix }}%
\providecommand \urlprefix  [0]{URL }%
\providecommand \Eprint [0]{\href }%
\providecommand \doibase [0]{http://dx.doi.org/}%
\providecommand \selectlanguage [0]{\@gobble}%
\providecommand \bibinfo  [0]{\@secondoftwo}%
\providecommand \bibfield  [0]{\@secondoftwo}%
\providecommand \translation [1]{[#1]}%
\providecommand \BibitemOpen [0]{}%
\providecommand \bibitemStop [0]{}%
\providecommand \bibitemNoStop [0]{.\EOS\space}%
\providecommand \EOS [0]{\spacefactor3000\relax}%
\providecommand \BibitemShut  [1]{\csname bibitem#1\endcsname}%
\let\auto@bib@innerbib\@empty
%</preamble>
\bibitem [{\citenamefont {Leinaas}\ and\ \citenamefont
  {Myrheim}(1977)}]{Leinaas77}%
  \BibitemOpen
  \bibfield  {author} {\bibinfo {author} {\bibfnamefont {J.~M.}\ \bibnamefont
  {Leinaas}}\ and\ \bibinfo {author} {\bibfnamefont {J.}~\bibnamefont
  {Myrheim}},\ }\bibfield  {title} {\enquote {\bibinfo {title} {{On the theory
  of identical particles}},}\ }\href {\doibase 10.1007/BF02727953} {\bibfield
  {journal} {\bibinfo  {journal} {Il Nuovo Cimento B}\ }\textbf {\bibinfo
  {volume} {37}},\ \bibinfo {pages} {1} (\bibinfo {year} {1977})}\BibitemShut
  {NoStop}%
\bibitem [{\citenamefont {Wilczek}(1982)}]{Wilczek82}%
  \BibitemOpen
  \bibfield  {author} {\bibinfo {author} {\bibfnamefont {F.}~\bibnamefont
  {Wilczek}},\ }\bibfield  {title} {\enquote {\bibinfo {title} {Quantum
  mechanics of fractional-spin particles},}\ }\href {\doibase
  10.1103/PhysRevLett.49.957} {\bibfield  {journal} {\bibinfo  {journal} {Phys.
  Rev. Lett.}\ }\textbf {\bibinfo {volume} {49}},\ \bibinfo {pages} {957}
  (\bibinfo {year} {1982})}\BibitemShut {NoStop}%
\bibitem [{\citenamefont {Kitaev}(2003)}]{Kitaev03}%
  \BibitemOpen
  \bibfield  {author} {\bibinfo {author} {\bibfnamefont {A.~Yu.}\ \bibnamefont
  {Kitaev}},\ }\bibfield  {title} {\enquote {\bibinfo {title} {{Fault-tolerant
  quantum computation by anyons}},}\ }\href {\doibase
  10.1016/S0003-4916(02)00018-0} {\bibfield  {journal} {\bibinfo  {journal}
  {Ann. Phys. (NY)}\ }\textbf {\bibinfo {volume} {303}},\ \bibinfo {pages} {2}
  (\bibinfo {year} {2003})}\BibitemShut {NoStop}%
\bibitem [{\citenamefont {Freedman}\ \emph {et~al.}(2003)\citenamefont
  {Freedman}, \citenamefont {Kitaev}, \citenamefont {Larsen},\ and\
  \citenamefont {Wang}}]{Freedman03}%
  \BibitemOpen
  \bibfield  {author} {\bibinfo {author} {\bibfnamefont {M.~H.}\ \bibnamefont
  {Freedman}}, \bibinfo {author} {\bibfnamefont {A.}~\bibnamefont {Kitaev}},
  \bibinfo {author} {\bibfnamefont {M.~J.}\ \bibnamefont {Larsen}}, \ and\
  \bibinfo {author} {\bibfnamefont {Z.}~\bibnamefont {Wang}},\ }\bibfield
  {title} {\enquote {\bibinfo {title} {{Topological quantum computation}},}\
  }\href {\doibase 10.1090/S0273-0979-02-00964-3} {\bibfield  {journal}
  {\bibinfo  {journal} {Bull. Am. Math. Soc.}\ }\textbf {\bibinfo {volume}
  {40}},\ \bibinfo {pages} {31} (\bibinfo {year} {2003})}\BibitemShut {NoStop}%
\bibitem [{Pre()}]{Preskill_HP}%
  \BibitemOpen
  \href@noop {} {}\bibinfo {note} {See
  {\href{http://www.theory.caltech.edu/people/preskill/ph219/}{http://www.theory.caltech.edu/people/preskill/ph219/}
  for a pedagogical introduction}}\BibitemShut {NoStop}%
\bibitem [{\citenamefont {Nayak}\ \emph {et~al.}(2008)\citenamefont {Nayak},
  \citenamefont {Simon}, \citenamefont {Stern},\ and\ \citenamefont
  {Freedman}}]{Nayak08}%
  \BibitemOpen
  \bibfield  {author} {\bibinfo {author} {\bibfnamefont {C.}~\bibnamefont
  {Nayak}}, \bibinfo {author} {\bibfnamefont {S.~H.}\ \bibnamefont {Simon}},
  \bibinfo {author} {\bibfnamefont {A.}~\bibnamefont {Stern}}, \ and\ \bibinfo
  {author} {\bibfnamefont {M.}~\bibnamefont {Freedman}},\ }\bibfield  {title}
  {\enquote {\bibinfo {title} {{Non-Abelian Anyons and Topological Quantum
  Computation}},}\ }\href {\doibase 10.1103/RevModPhys.80.1083} {\bibfield
  {journal} {\bibinfo  {journal} {Rev. Mod. Phys.}\ }\textbf {\bibinfo {volume}
  {80}},\ \bibinfo {pages} {1083} (\bibinfo {year} {2008})}\BibitemShut
  {NoStop}%
\bibitem [{Wan()}]{Wang_book}%
  \BibitemOpen
  \href {\doibase 10.1090/cbms/112} {}\bibinfo {note} {{Z. Wang, {\it
  Topological Quantum Computation}, CBMS Regional Conference Series in
  Mathematics, No. 112 (American Mathematical Society, Providence,
  2010)}}\BibitemShut {NoStop}%
\bibitem [{\citenamefont {Moore}\ and\ \citenamefont {Read}(1991)}]{Moore91}%
  \BibitemOpen
  \bibfield  {author} {\bibinfo {author} {\bibfnamefont {G.}~\bibnamefont
  {Moore}}\ and\ \bibinfo {author} {\bibfnamefont {N.}~\bibnamefont {Read}},\
  }\bibfield  {title} {\enquote {\bibinfo {title} {{Nonabelions in the
  fractional quantum Hall effect}},}\ }\href {\doibase
  10.1016/0550-3213(91)90407-O} {\bibfield  {journal} {\bibinfo  {journal}
  {Nucl. Phys. B}\ }\textbf {\bibinfo {volume} {360}},\ \bibinfo {pages} {362}
  (\bibinfo {year} {1991})}\BibitemShut {NoStop}%
\bibitem [{\citenamefont {Wen}(1991)}]{Wen91}%
  \BibitemOpen
  \bibfield  {author} {\bibinfo {author} {\bibfnamefont {X.-G.}\ \bibnamefont
  {Wen}},\ }\bibfield  {title} {\enquote {\bibinfo {title} {{Non-Abelian
  statistics in the fractional quantum Hall states}},}\ }\href {\doibase
  10.1103/PhysRevLett.66.802} {\bibfield  {journal} {\bibinfo  {journal} {Phys.
  Rev. Lett.}\ }\textbf {\bibinfo {volume} {66}},\ \bibinfo {pages} {802}
  (\bibinfo {year} {1991})}\BibitemShut {NoStop}%
\bibitem [{\citenamefont {Wen}(2013)}]{Wen13}%
  \BibitemOpen
  \bibfield  {author} {\bibinfo {author} {\bibfnamefont {X.-G.}\ \bibnamefont
  {Wen}},\ }\bibfield  {title} {\enquote {\bibinfo {title} {{Topological order:
  from long-range entangled quantum matter to a unification of light and
  electrons}},}\ }\href {\doibase 10.1155/2013/198710} {\bibfield  {journal}
  {\bibinfo  {journal} {ISRN Condens. Matter Phys.}\ }\textbf {\bibinfo
  {volume} {2013}},\ \bibinfo {pages} {198710} (\bibinfo {year}
  {2013})}\BibitemShut {NoStop}%
\bibitem [{\citenamefont {Rowell}\ \emph {et~al.}(2009)\citenamefont {Rowell},
  \citenamefont {Stong},\ and\ \citenamefont {Wang}}]{Rowell09}%
  \BibitemOpen
  \bibfield  {author} {\bibinfo {author} {\bibfnamefont {E.}~\bibnamefont
  {Rowell}}, \bibinfo {author} {\bibfnamefont {R.}~\bibnamefont {Stong}}, \
  and\ \bibinfo {author} {\bibfnamefont {Z.}~\bibnamefont {Wang}},\ }\bibfield
  {title} {\enquote {\bibinfo {title} {{On Classification of Modular Tensor
  Categories}},}\ }\href {\doibase 10.1007/s00220-009-0908-z} {\bibfield
  {journal} {\bibinfo  {journal} {Commun. Math. Phys.}\ }\textbf {\bibinfo
  {volume} {292}},\ \bibinfo {pages} {343} (\bibinfo {year}
  {2009})}\BibitemShut {NoStop}%
\bibitem [{\citenamefont {Wen}(2016)}]{Wen16}%
  \BibitemOpen
  \bibfield  {author} {\bibinfo {author} {\bibfnamefont {X.-G.}\ \bibnamefont
  {Wen}},\ }\bibfield  {title} {\enquote {\bibinfo {title} {{A theory of 2+1D
  bosonic topological orders}},}\ }\href {\doibase 10.1093/nsr/nwv077}
  {\bibfield  {journal} {\bibinfo  {journal} {Natl. Sci. Rev.}\ }\textbf
  {\bibinfo {volume} {3}},\ \bibinfo {pages} {68} (\bibinfo {year}
  {2016})}\BibitemShut {NoStop}%
\bibitem [{\citenamefont {Schoutens}\ and\ \citenamefont
  {Wen}(2016)}]{Schoutens16}%
  \BibitemOpen
  \bibfield  {author} {\bibinfo {author} {\bibfnamefont {K.}~\bibnamefont
  {Schoutens}}\ and\ \bibinfo {author} {\bibfnamefont {X.-G.}\ \bibnamefont
  {Wen}},\ }\bibfield  {title} {\enquote {\bibinfo {title} {{Simple-current
  algebra constructions of 2+1-dimensional topological orders}},}\ }\href
  {\doibase 10.1103/PhysRevB.93.045109} {\bibfield  {journal} {\bibinfo
  {journal} {Phys. Rev. B}\ }\textbf {\bibinfo {volume} {93}},\ \bibinfo
  {pages} {045109} (\bibinfo {year} {2016})}\BibitemShut {NoStop}%
\bibitem [{\citenamefont {Levin}\ and\ \citenamefont {Wen}(2005)}]{Levin05}%
  \BibitemOpen
  \bibfield  {author} {\bibinfo {author} {\bibfnamefont {M.~A.}\ \bibnamefont
  {Levin}}\ and\ \bibinfo {author} {\bibfnamefont {X.-G.}\ \bibnamefont
  {Wen}},\ }\bibfield  {title} {\enquote {\bibinfo {title} {{String-net
  condensation: A physical mechanism for topological phases}},}\ }\href
  {\doibase 10.1103/PhysRevB.71.045110} {\bibfield  {journal} {\bibinfo
  {journal} {Phys. Rev. B}\ }\textbf {\bibinfo {volume} {71}},\ \bibinfo
  {pages} {045110} (\bibinfo {year} {2005})}\BibitemShut {NoStop}%
\bibitem [{\citenamefont {Lin}\ and\ \citenamefont {Levin}(2014)}]{Lin14}%
  \BibitemOpen
  \bibfield  {author} {\bibinfo {author} {\bibfnamefont {C.-H.}\ \bibnamefont
  {Lin}}\ and\ \bibinfo {author} {\bibfnamefont {M.}~\bibnamefont {Levin}},\
  }\bibfield  {title} {\enquote {\bibinfo {title} {{Generalizations and
  limitations of string-net models}},}\ }\href {\doibase
  10.1103/PhysRevB.89.195130} {\bibfield  {journal} {\bibinfo  {journal} {Phys.
  Rev. B}\ }\textbf {\bibinfo {volume} {89}},\ \bibinfo {pages} {195130}
  (\bibinfo {year} {2014})}\BibitemShut {NoStop}%
\bibitem [{\citenamefont {Lake}\ and\ \citenamefont {Wu}(2016)}]{Lake16}%
  \BibitemOpen
  \bibfield  {author} {\bibinfo {author} {\bibfnamefont {E.}~\bibnamefont
  {Lake}}\ and\ \bibinfo {author} {\bibfnamefont {Y.-S.}\ \bibnamefont {Wu}},\
  }\bibfield  {title} {\enquote {\bibinfo {title} {{Signatures of broken parity
  and time-reversal symmetry in generalized string-net models}},}\ }\href
  {\doibase 10.1103/PhysRevB.94.115139} {\bibfield  {journal} {\bibinfo
  {journal} {Phys. Rev. B}\ }\textbf {\bibinfo {volume} {94}},\ \bibinfo
  {pages} {115139} (\bibinfo {year} {2016})}\BibitemShut {NoStop}%
\bibitem [{\citenamefont {Gils}\ \emph {et~al.}(2009)\citenamefont {Gils},
  \citenamefont {Trebst}, \citenamefont {Kitaev}, \citenamefont {Ludwig},
  \citenamefont {Troyer},\ and\ \citenamefont {Wang}}]{Gils09_1}%
  \BibitemOpen
  \bibfield  {author} {\bibinfo {author} {\bibfnamefont {C.}~\bibnamefont
  {Gils}}, \bibinfo {author} {\bibfnamefont {S.}~\bibnamefont {Trebst}},
  \bibinfo {author} {\bibfnamefont {A.}~\bibnamefont {Kitaev}}, \bibinfo
  {author} {\bibfnamefont {A.~W.~W.}\ \bibnamefont {Ludwig}}, \bibinfo {author}
  {\bibfnamefont {M.}~\bibnamefont {Troyer}}, \ and\ \bibinfo {author}
  {\bibfnamefont {Z.}~\bibnamefont {Wang}},\ }\bibfield  {title} {\enquote
  {\bibinfo {title} {{Topology-driven quantum phase transitions in
  time-reversal-invariant anyonic quantum liquids}},}\ }\href {\doibase
  10.1038/nphys1396} {\bibfield  {journal} {\bibinfo  {journal} {Nat. Phys.}\
  }\textbf {\bibinfo {volume} {5}},\ \bibinfo {pages} {834} (\bibinfo {year}
  {2009})}\BibitemShut {NoStop}%
\bibitem [{\citenamefont {Gils}()}]{Gils09_3}%
  \BibitemOpen
  \bibfield  {author} {\bibinfo {author} {\bibfnamefont {C.}~\bibnamefont
  {Gils}},\ }\href@noop {} {\enquote {\bibinfo {title} {{Ashkin-Teller
  universality in a quantum double model of Ising anyons}},}\ }\bibinfo {note}
  {\href{http://iopscience.iop.org/1742-5468/2009/07/P07019/}{J. Stat. Mech.,
  P07019 (2009)}}\BibitemShut {NoStop}%
\bibitem [{\citenamefont {Ardonne}\ \emph {et~al.}(2011)\citenamefont
  {Ardonne}, \citenamefont {Gukelberger}, \citenamefont {Ludwig}, \citenamefont
  {Trebst},\ and\ \citenamefont {Troyer}}]{Ardonne11}%
  \BibitemOpen
  \bibfield  {author} {\bibinfo {author} {\bibfnamefont {E.}~\bibnamefont
  {Ardonne}}, \bibinfo {author} {\bibfnamefont {J.}~\bibnamefont
  {Gukelberger}}, \bibinfo {author} {\bibfnamefont {A.~W.~W.}\ \bibnamefont
  {Ludwig}}, \bibinfo {author} {\bibfnamefont {S.}~\bibnamefont {Trebst}}, \
  and\ \bibinfo {author} {\bibfnamefont {M.}~\bibnamefont {Troyer}},\
  }\bibfield  {title} {\enquote {\bibinfo {title} {{Microscopic models of
  interacting Yang-Lee anyons}},}\ }\href {\doibase
  10.1088/1367-2630/13/4/045006} {\bibfield  {journal} {\bibinfo  {journal}
  {New J. Phys.}\ }\textbf {\bibinfo {volume} {13}},\ \bibinfo {pages} {045006}
  (\bibinfo {year} {2011})}\BibitemShut {NoStop}%
\bibitem [{\citenamefont {Burnell}\ \emph {et~al.}(2011)\citenamefont
  {Burnell}, \citenamefont {Simon},\ and\ \citenamefont
  {Slingerland}}]{Burnell11_2}%
  \BibitemOpen
  \bibfield  {author} {\bibinfo {author} {\bibfnamefont {F.~J.}\ \bibnamefont
  {Burnell}}, \bibinfo {author} {\bibfnamefont {S.~H.}\ \bibnamefont {Simon}},
  \ and\ \bibinfo {author} {\bibfnamefont {J.~K.}\ \bibnamefont
  {Slingerland}},\ }\bibfield  {title} {\enquote {\bibinfo {title}
  {{Condensation of achiral simple currents in topological lattice models:
  Hamiltonian study of topological symmetry breaking}},}\ }\href {\doibase
  10.1103/PhysRevB.84.125434} {\bibfield  {journal} {\bibinfo  {journal} {Phys.
  Rev. B}\ }\textbf {\bibinfo {volume} {84}},\ \bibinfo {pages} {125434}
  (\bibinfo {year} {2011})}\BibitemShut {NoStop}%
\bibitem [{\citenamefont {Schulz}\ \emph {et~al.}(2013)\citenamefont {Schulz},
  \citenamefont {Dusuel}, \citenamefont {Schmidt},\ and\ \citenamefont
  {Vidal}}]{Schulz13}%
  \BibitemOpen
  \bibfield  {author} {\bibinfo {author} {\bibfnamefont {M.~D.}\ \bibnamefont
  {Schulz}}, \bibinfo {author} {\bibfnamefont {S.}~\bibnamefont {Dusuel}},
  \bibinfo {author} {\bibfnamefont {K.~P.}\ \bibnamefont {Schmidt}}, \ and\
  \bibinfo {author} {\bibfnamefont {J.}~\bibnamefont {Vidal}},\ }\bibfield
  {title} {\enquote {\bibinfo {title} {{Topological Phase Transitions in the
  Golden String-Net Model}},}\ }\href {\doibase 10.1103/PhysRevLett.110.147203}
  {\bibfield  {journal} {\bibinfo  {journal} {Phys. Rev. Lett.}\ }\textbf
  {\bibinfo {volume} {110}},\ \bibinfo {pages} {147203} (\bibinfo {year}
  {2013})}\BibitemShut {NoStop}%
\bibitem [{\citenamefont {Schulz}\ \emph {et~al.}(2014)\citenamefont {Schulz},
  \citenamefont {Dusuel}, \citenamefont {Misguich}, \citenamefont {Schmidt},\
  and\ \citenamefont {Vidal}}]{Schulz14}%
  \BibitemOpen
  \bibfield  {author} {\bibinfo {author} {\bibfnamefont {M.~D.}\ \bibnamefont
  {Schulz}}, \bibinfo {author} {\bibfnamefont {S.}~\bibnamefont {Dusuel}},
  \bibinfo {author} {\bibfnamefont {G.}~\bibnamefont {Misguich}}, \bibinfo
  {author} {\bibfnamefont {K.~P.}\ \bibnamefont {Schmidt}}, \ and\ \bibinfo
  {author} {\bibfnamefont {J.}~\bibnamefont {Vidal}},\ }\bibfield  {title}
  {\enquote {\bibinfo {title} {{Ising anyons with a string tension}},}\ }\href
  {\doibase 10.1103/PhysRevB.89.201103} {\bibfield  {journal} {\bibinfo
  {journal} {Phys. Rev. B}\ }\textbf {\bibinfo {volume} {89}},\ \bibinfo
  {pages} {201103} (\bibinfo {year} {2014})}\BibitemShut {NoStop}%
\bibitem [{\citenamefont {Schulz}\ \emph {et~al.}(2015)\citenamefont {Schulz},
  \citenamefont {Dusuel},\ and\ \citenamefont {Vidal}}]{Schulz15}%
  \BibitemOpen
  \bibfield  {author} {\bibinfo {author} {\bibfnamefont {M.~D.}\ \bibnamefont
  {Schulz}}, \bibinfo {author} {\bibfnamefont {S.}~\bibnamefont {Dusuel}}, \
  and\ \bibinfo {author} {\bibfnamefont {J.}~\bibnamefont {Vidal}},\ }\bibfield
   {title} {\enquote {\bibinfo {title} {{Russian doll spectrum in a non-Abelian
  string-net ladder}},}\ }\href {\doibase 10.1103/PhysRevB.91.155110}
  {\bibfield  {journal} {\bibinfo  {journal} {Phys. Rev. B}\ }\textbf {\bibinfo
  {volume} {91}},\ \bibinfo {pages} {155110} (\bibinfo {year}
  {2015})}\BibitemShut {NoStop}%
\bibitem [{\citenamefont {Dusuel}\ and\ \citenamefont
  {Vidal}(2015)}]{Dusuel15}%
  \BibitemOpen
  \bibfield  {author} {\bibinfo {author} {\bibfnamefont {S.}~\bibnamefont
  {Dusuel}}\ and\ \bibinfo {author} {\bibfnamefont {J.}~\bibnamefont {Vidal}},\
  }\bibfield  {title} {\enquote {\bibinfo {title} {{Mean-field ansatz for
  topological phases with string tension}},}\ }\href {\doibase
  10.1103/PhysRevB.92.125150} {\bibfield  {journal} {\bibinfo  {journal} {Phys.
  Rev. B}\ }\textbf {\bibinfo {volume} {92}},\ \bibinfo {pages} {125150}
  (\bibinfo {year} {2015})}\BibitemShut {NoStop}%
\bibitem [{\citenamefont {Schulz}\ and\ \citenamefont
  {Burnell}(2016)}]{Schulz16}%
  \BibitemOpen
  \bibfield  {author} {\bibinfo {author} {\bibfnamefont {M.~D.}\ \bibnamefont
  {Schulz}}\ and\ \bibinfo {author} {\bibfnamefont {F.~J.}\ \bibnamefont
  {Burnell}},\ }\bibfield  {title} {\enquote {\bibinfo {title} {{Frustrated
  topological symmetry breaking: geometrical frustration and anyon
  condensation}},}\ }\href {\doibase 10.1103/PhysRevB.94.165110} {\bibfield
  {journal} {\bibinfo  {journal} {Phys. Rev. B}\ }\textbf {\bibinfo {volume}
  {94}},\ \bibinfo {pages} {165110} (\bibinfo {year} {2016})}\BibitemShut
  {NoStop}%
\bibitem [{Note1()}]{Note1}%
  \BibitemOpen
  \bibinfo {note} {For non-Abelian theories and for $g>0$, one may generate
  single-flux excitations by using nonlocal operators.}\BibitemShut {Stop}%
\bibitem [{\citenamefont {Dorey}\ \emph
  {et~al.}(2003{\natexlab{a}})\citenamefont {Dorey}, \citenamefont
  {Pocklington},\ and\ \citenamefont {Tateo}}]{Dorey03_1}%
  \BibitemOpen
  \bibfield  {author} {\bibinfo {author} {\bibfnamefont {P.}~\bibnamefont
  {Dorey}}, \bibinfo {author} {\bibfnamefont {A.}~\bibnamefont {Pocklington}},
  \ and\ \bibinfo {author} {\bibfnamefont {R.}~\bibnamefont {Tateo}},\
  }\bibfield  {title} {\enquote {\bibinfo {title} {{Integrable aspects of the
  scaling q-state Potts models I: bound states and bootstrap closure}},}\
  }\href {\doibase 10.1016/S0550-3213(03)00181-0} {\bibfield  {journal}
  {\bibinfo  {journal} {Nucl. Phys. B}\ }\textbf {\bibinfo {volume} {661}},\
  \bibinfo {pages} {425} (\bibinfo {year} {2003}{\natexlab{a}})}\BibitemShut
  {NoStop}%
\bibitem [{\citenamefont {Dorey}\ \emph
  {et~al.}(2003{\natexlab{b}})\citenamefont {Dorey}, \citenamefont
  {Pocklington},\ and\ \citenamefont {Tateo}}]{Dorey03_2}%
  \BibitemOpen
  \bibfield  {author} {\bibinfo {author} {\bibfnamefont {P.}~\bibnamefont
  {Dorey}}, \bibinfo {author} {\bibfnamefont {A.}~\bibnamefont {Pocklington}},
  \ and\ \bibinfo {author} {\bibfnamefont {R.}~\bibnamefont {Tateo}},\
  }\bibfield  {title} {\enquote {\bibinfo {title} {{Integrable aspects of the
  scaling q-state Potts models II: finite-size effects}},}\ }\href {\doibase
  10.1016/S0550-3213(03)00182-2} {\bibfield  {journal} {\bibinfo  {journal}
  {Nucl. Phys. B}\ }\textbf {\bibinfo {volume} {661}},\ \bibinfo {pages} {464}
  (\bibinfo {year} {2003}{\natexlab{b}})}\BibitemShut {NoStop}%
\bibitem [{\citenamefont {He}\ \emph {et~al.}(1990)\citenamefont {He},
  \citenamefont {Hamer},\ and\ \citenamefont {Oitmaa}}]{He90}%
  \BibitemOpen
  \bibfield  {author} {\bibinfo {author} {\bibfnamefont {H.-X.}\ \bibnamefont
  {He}}, \bibinfo {author} {\bibfnamefont {C.~J.}\ \bibnamefont {Hamer}}, \
  and\ \bibinfo {author} {\bibfnamefont {J.}~\bibnamefont {Oitmaa}},\
  }\bibfield  {title} {\enquote {\bibinfo {title} {{High-temperature series
  expansion for the (2+1)-dimensional Ising model}},}\ }\href {\doibase
  10.1088/0305-4470/23/10/018} {\bibfield  {journal} {\bibinfo  {journal} {J.
  Phys. A}\ }\textbf {\bibinfo {volume} {23}},\ \bibinfo {pages} {1775}
  (\bibinfo {year} {1990})}\BibitemShut {NoStop}%
\bibitem [{\citenamefont {Hamer}\ \emph {et~al.}(1992)\citenamefont {Hamer},
  \citenamefont {Oitmaa},\ and\ \citenamefont {Weihong}}]{Hamer92}%
  \BibitemOpen
  \bibfield  {author} {\bibinfo {author} {\bibfnamefont {C.~J.}\ \bibnamefont
  {Hamer}}, \bibinfo {author} {\bibfnamefont {J.}~\bibnamefont {Oitmaa}}, \
  and\ \bibinfo {author} {\bibfnamefont {Z.}~\bibnamefont {Weihong}},\
  }\bibfield  {title} {\enquote {\bibinfo {title} {{Series analysis of the
  3-state Potts model in (2+1) dimensions}},}\ }\href {\doibase
  10.1088/0305-4470/25/7/023} {\bibfield  {journal} {\bibinfo  {journal} {J.
  Phys. A}\ }\textbf {\bibinfo {volume} {25}},\ \bibinfo {pages} {1821}
  (\bibinfo {year} {1992})}\BibitemShut {NoStop}%
\bibitem [{\citenamefont {Kamfor}\ \emph {et~al.}(2014)\citenamefont {Kamfor},
  \citenamefont {Dusuel}, \citenamefont {Vidal},\ and\ \citenamefont
  {Schmidt}}]{Kamfor14}%
  \BibitemOpen
  \bibfield  {author} {\bibinfo {author} {\bibfnamefont {M.}~\bibnamefont
  {Kamfor}}, \bibinfo {author} {\bibfnamefont {S.}~\bibnamefont {Dusuel}},
  \bibinfo {author} {\bibfnamefont {J.}~\bibnamefont {Vidal}}, \ and\ \bibinfo
  {author} {\bibfnamefont {K.~P.}\ \bibnamefont {Schmidt}},\ }\bibfield
  {title} {\enquote {\bibinfo {title} {{Spectroscopy of a topological
  phase}},}\ }\href {\doibase 10.1103/PhysRevB.89.045411} {\bibfield  {journal}
  {\bibinfo  {journal} {Phys. Rev. B}\ }\textbf {\bibinfo {volume} {89}},\
  \bibinfo {pages} {045411} (\bibinfo {year} {2014})}\BibitemShut {NoStop}%
\bibitem [{\citenamefont {Zamolodchikov}(1989)}]{Zamolodchikov89}%
  \BibitemOpen
  \bibfield  {author} {\bibinfo {author} {\bibfnamefont {A.~B.}\ \bibnamefont
  {Zamolodchikov}},\ }\bibfield  {title} {\enquote {\bibinfo {title}
  {{Integrals of motion and $S$-matrix of the (scaled) $T=T_{\rm c}$ Ising
  model with magnetic field}},}\ }\href {\doibase 10.1142/S0217751X8900176X}
  {\bibfield  {journal} {\bibinfo  {journal} {Int. J. Mod. Phys. B}\ }\textbf
  {\bibinfo {volume} {04}},\ \bibinfo {pages} {4235} (\bibinfo {year}
  {1989})}\BibitemShut {NoStop}%
\bibitem [{\citenamefont {Coldea}\ \emph {et~al.}(2010)\citenamefont {Coldea},
  \citenamefont {Tennant}, \citenamefont {Wheeler}, \citenamefont {Wawrzynska},
  \citenamefont {Prabhakaran}, \citenamefont {Telling}, \citenamefont
  {Habichtand}, \citenamefont {Smeibidl},\ and\ \citenamefont
  {Kiefer}}]{Coldea10}%
  \BibitemOpen
  \bibfield  {author} {\bibinfo {author} {\bibfnamefont {R.}~\bibnamefont
  {Coldea}}, \bibinfo {author} {\bibfnamefont {D.~A.}\ \bibnamefont {Tennant}},
  \bibinfo {author} {\bibfnamefont {E.~M.}\ \bibnamefont {Wheeler}}, \bibinfo
  {author} {\bibfnamefont {E.}~\bibnamefont {Wawrzynska}}, \bibinfo {author}
  {\bibfnamefont {D.}~\bibnamefont {Prabhakaran}}, \bibinfo {author}
  {\bibfnamefont {M.}~\bibnamefont {Telling}}, \bibinfo {author} {\bibfnamefont
  {K.}~\bibnamefont {Habichtand}}, \bibinfo {author} {\bibfnamefont
  {P.}~\bibnamefont {Smeibidl}}, \ and\ \bibinfo {author} {\bibfnamefont
  {K.}~\bibnamefont {Kiefer}},\ }\bibfield  {title} {\enquote {\bibinfo {title}
  {{Quantum Criticality in an Ising Chain: Experimental Evidence for Emergent
  $E_8$ Symmetry}},}\ }\href {\doibase 10.1126/science.1180085} {\bibfield
  {journal} {\bibinfo  {journal} {Science}\ }\textbf {\bibinfo {volume}
  {327}},\ \bibinfo {pages} {177} (\bibinfo {year} {2010})}\BibitemShut
  {NoStop}%
\end{thebibliography}


\end{document}